\documentclass[
preprint,
aps,prl,showpacs,preprintnumbers,amsmath,amssymb]{revtex4}
\usepackage[english]{babel}
\usepackage[T1]{fontenc}
\usepackage[utf8]{inputenc}
\usepackage{graphicx}
\usepackage{dcolumn}
\usepackage{amsfonts}
\usepackage{amsmath}
\usepackage{bm}
\usepackage[colorlinks=true, pdfstartview=FitV, linkcolor=blue, citecolor=blue, urlcolor=blue]{hyperref}

\begin{document}

\title{Reversal mechanism of an individual Ni nanotube simultaneously
  studied by torque and SQUID magnetometry}

\author{A. Buchter$^1$, J. Nagel$^2$, D. R\"{u}ffer$^3$, F. Xue$^1$,
  D.~P.~Weber$^1$, O. F. Kieler$^4$, T. Weimann$^4$, J. Kohlmann$^4$,
  A. B. Zorin$^4$, E. Russo-Averchi$^3$, R. Huber$^5$, P. Berberich$^5$,
  A. Fontcuberta i Morral$^3$, M. Kemmler$^2$, R. Kleiner$^2$,
  D. Koelle$^2$, D. Grundler$^{5,6}$} \email{grundler@ph.tum.de}
\author{M. Poggio$^1$} \email{martino.poggio@unibas.ch}

\affiliation{$^1$Department of Physics, University of Basel, 4056
  Basel, Switzerland; \\$^2$Physikalisches Institut and Center for
  Collective Quantum Phenomena in LISA$^{+}$, Universit\"{a}t
  T\"{u}bingen, 72076 T\"{u}bingen, Germany; \\$^3$Laboratoire des
  Mat\'{e}riaux Semiconducteurs, Institut des Mat\'{e}riaux, Ecole
  Polytechnique F\'{e}d\'{e}rale de Lausanne, 1015 Lausanne,
  Switzerland; \\$^4$Fachbereich 2.4 ``Quantenelektronik'',
  Physikalisch-Technische Bundesanstalt, 38116 Braunschweig, Germany;
  \\$^5$Lehrstuhl f\"{u}r Physik funktionaler Schichtsysteme, Physik
  Department E10, Technische Universit\"{a}t M\"{u}nchen, 85747
  Garching, Germany; \\$^6$Facult\'{e}
  Sciences et Technique de l'Ing\'{e}nieur, Ecole Polytechnique
  F\'{e}d\'{e}rale de Lausanne, 1015 Lausanne, Switzerland}

\date{\today}


\begin{abstract}
  Using an optimally coupled nanometer-scale SQUID, we measure the
  magnetic flux originating from an individual ferromagnetic Ni
  nanotube attached to a Si cantilever.  At the same time, we detect
  the nanotube's volume magnetization using torque magnetometry.  We
  observe both the predicted reversible and irreversible reversal
  processes.  A detailed comparison with micromagnetic simulations
  suggests that vortex-like states are formed in different segments of
  the individual nanotube.  Such stray-field free states are
  interesting for memory applications and non-invasive sensing.
\end{abstract}


\maketitle

\begin{figure}[t]
	\includegraphics[]{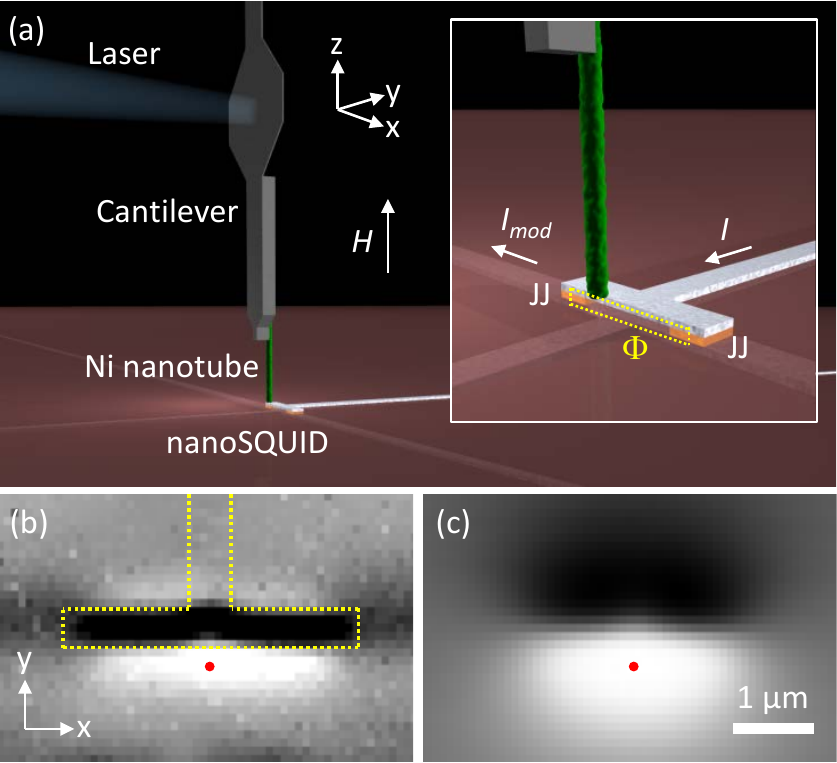}
	\caption{\label{f:setup} (Color) (a) Sketch of the apparatus
          (inset: zoomed-in view; dashed line indicates SQUID loop).
          Gray-scale maps of (b) $\Delta f (x, y)$ and (c) $\Phi(x,
          y)$ taken simultaneously at a distance $z = 280$ nm with
          $H = 0$. $\Delta f$ ($\Phi$) ranges from -$170$ to $430$~Hz
          ($-0.08~\Phi_0$ to $0.08~\Phi_0$). Dashed lines indicate the
          T-shaped SQUID arm and dots the operating position.}
\end{figure}

Recent experimental and theoretical work has demonstrated that
nanometer-scale magnets, as a result of their low-dimensionality,
display magnetic configurations not present in their macroscopic
counterparts \cite{Wang:2002,Streubel:2012,StreubelAPL:2012}.  Such
work is driven by both fundamental questions about nanometer-scale
magnetism and the potential for applying nanomagnets as elements in
high-density memories \cite{Parkin:2008}, in high-resolution imaging
\cite{Khizroev:2002,Poggio:2010,Campanella:2011}, or as magnetic
sensors \cite{Maqableh:2012}.  Compared to nanowires, ferromagnetic
nanotubes are particularly interesting for magnetization reversal as
they avoid the Bloch point structure \cite{Hertel:2004}.  Different
reversal processes via curling, vortex wall formation, and propagation
have been predicted
\cite{LanderosAPL:2007,Escrig:2008,Landeros:2009,Landeros:2010}.  Due
to their inherently small magnetic moment, experimental investigations
have often been conducted on large ensembles.  The results, however,
are difficult to interpret due to stray-field interactions and the
distribution in size and orientation of the individual nanotubes
\cite{BachmannJACS:2007,DaubJAP:2007,EscrigAPL:2008,Escrig:2008,BachmannJAP:2009,Albrecht:2011}.
In a pioneering work, Wernsdorfer {\em et al.} \cite{Wernsdorfer:1996}
investigated the magnetic reversal of an individual Ni nanowire at 4~K
using a miniaturized SQUID.  Detecting the stray magnetic flux $\Phi$
from one end of the nanowire as a function of magnetic field
$\mathbf{H}$, $\Phi$ was assumed to be approximately proportional to
the projection of the total magnetization $\mathbf{M}$ along the
nanowire axis.  At the time, $M(H)$ of the individual nanowire was not
accessible and micromagnetic simulations were conducted only a decade
later \cite{Hertel:2004}.  Here we present a technique to
simultaneously measure $\Phi(H)$ and $M(H)$ of a single
low-dimensional magnet.  Using a scanning nanoSQUID and a
cantilever-based torque magnetometer (Fig.~\ref{f:setup})
\cite{Nagel:2013}, we investigate a Ni nanotube producing $\Phi(H)$
with a nearly square hysteresis, similar to the Ni nanowire of
Ref.~\cite{Wernsdorfer:1996}.  $M(H)$, however, displays a more
complex behavior composed of reversible and irreversible
contributions, which we interpret in detail with micromagnetic
simulations.  In contrast to theoretical predictions, the experiment
suggests that magnetization reversal is not initiated from both ends.
If nanomagnets are to be optimized for storage or sensing
applications, such detailed investigations of nanoscale properties are
essential.

We use a direct current nanoSQUID formed by a loop containing two
superconductor-normal-superconductor Josephson junctions (JJs)
\cite{NagelAPL:2011,Wolbing:2013,SupMat} (Fig.\ \ref{f:setup} (a)).
Two T-shaped superconducting Nb arms are sputtered on top of each
other separated by an insulating layer of SiO$_2$.  The Nb arms are
connected via two planar 225-nm-thick Nb/HfTi/Nb JJs each with an area
of $200\times200$ nm$^2$.  These JJs and the $1.8$-$\mu$m-long Nb
leads form a SQUID loop in the $xz$-plane (shown in yellow in
Fig.~\ref{f:setup} (a)), through which we measure $\Phi$.  Atomic
layer deposition of Ni is used to prepare the nanotube around a GaAs
nanowire template grown by molecular beam epitaxy (MBE)
\cite{Ruffer:2012,Weber:2012}.  The GaAs core supports the structure,
making it mechanically robust.  The polycrystalline nanotube, which
does not exhibit magneto-crystalline anisotropy, has a $140 \pm 20$-nm
outer diameter, a $70 \pm 10$-nm inner diameter, and a $6.0 \pm
0.5$-$\mu$m length.  The error in the diameters results from the
roughness of the Ni film \cite{SupMat}.  The Ni nanotube is affixed to
the end of an ultrasoft Si cantilever \cite{Weber:2012}, such that it
protrudes from the tip by $4\,\mathrm{\mu m}$.  The cantilever is
120-$\mu$m-long, 4-$\mu$m-wide and 0.1-$\mu$m-thick.  It hangs above
the nanoSQUID in the pendulum geometry, inside a vacuum chamber
(pressure $< 10^{-6}$ mbar) at the bottom of a cryostat.  A 3D
piezo-electric positioning stage moves the nanoSQUID relative to the
Ni nanotube and an optical fiber interferometer is used to detect
deflections of the cantilever along $\hat{y}$ \cite{Rugar:1989}.  Fast
and accurate measurement of the cantilever's fundamental resonance
frequency $f_c$ is realized by self-oscillation at a fixed amplitude.
An external field $\mu_0\mathbf{H}$ of up to $2.8$~T can be applied
along the cantilever axis $\hat{z}$ using a superconducting magnet.
At $4.3\,\mathrm{K}$ and $\mu_0 H=0$, the cantilever, loaded with the
Ni nanotube and far from any surfaces, has an intrinsic resonance
frequency $f_c=f_0=3413\,\mathrm{Hz}$, a quality factor
$Q=Q_0=3.4\times10^{4}$, and spring constant of $k_0=90 \pm
10\,\mathrm{\mu N/m}$.  The magnetic flux due the Ni nanotube
$\Phi_{\text{NN}}(H)$ is evaluated from $\Phi_{\text{NN}}(H) = \Phi(H)
- \Phi_{\text{ref}}(H)$, where the flux $\Phi(H)$ is measured with the
nanotube close to the nanoSQUID, while $\Phi_{\text{ref}}(H)$ is
measured with the nanotube several $\mu$m away such that the stray
flux is negligible.  Therefore $\Phi_{\text{ref}}(H) \propto H$, due
to the small fraction of $\mathbf{H}$ that couples through the
nanoSQUID given its imperfect alignment with $\hat{z}$.  Once
calibrated, we also use $\Phi_{\text{ref}}(H)$ to measure the $\mu_0
H$ axis of our plots, removing effects due to hysteresis in the
superconducting magnet.  Such a field calibration was not possible for
the integrated SQUID of Ref.~\cite{Wernsdorfer:1996}.  We also perform
dynamic-mode cantilever magnetometry \cite{Stipe:2001.2}, which is
sensitive to the dynamic component of the magnetic torque acting
between $\mathbf{H}$ and the magnetization $\mathbf{M}$ of the Ni
nanotube.  In order to extract $M(H)$, we measure the field-dependent
frequency shift $\Delta f (H) = f_c (H) - f_0$.  Micromagnetic
simulations are performed with NMAG \cite{NMAG} which provides
finite-element modeling by adapting a mesh to the curved inner and
outer surfaces of the nanotube.  We simulate 30-nm thick nanotubes of
different lengths $l$ and the same 70-nm inner diameter.  We assume
magnetically isotropic Ni consistent with earlier studies
\cite{Ruffer:2012}, a saturation magnetization $M_{\text{S}} = 406$
kA/m \cite{Kittel:2005}, and exchange coupling constant of $7 \times
10^{12}$ J/m \cite{Knittel:2011}.

\begin{figure}[t]
  \includegraphics[]{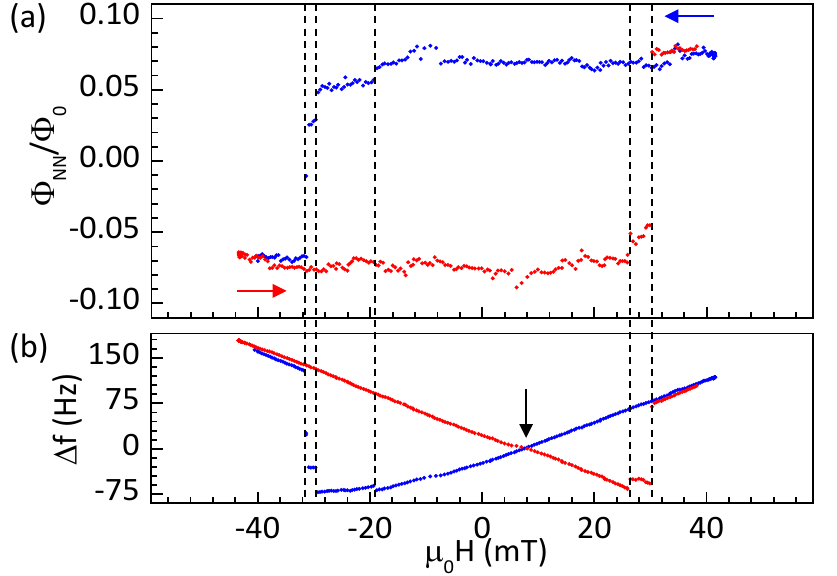}
  \caption{\label{f:SQUIDHysteresis} (Color) Simultaneously measured
    hysteresis loops of (a) $\Phi_{\text{NN}} (H)$ and (b) $\Delta f
    (H)$ at $z=450$ nm.  Red (blue) points represent data taken while
    sweeping $H$ in the positive (negative) direction.  Dashed lines
    indicate discontinuities (magnetic switching fields $H_{\rm
      sw,e}$) appearing in both $\Phi_{\text{NN}} (H)$ and $\Delta f
    (H)$.}
\end{figure}

We first scan the nanoSQUID under the cantilever with attached Ni
nanotube, to map the coupling between them.  To ensure that the scan
is done with the nanotube in a well-defined magnetic state, we first
saturate it along its easy axis ($\hat{z}$).  Scans are then made at
$H = 0$ in the $xy$-plane at a fixed height $z$, i.e.\ for a fixed
distance between the top of the SQUID device and the bottom end of the
Ni nanotube.  $\Delta f(x, y) = f_c(x, y) - f_0$ and $\Phi(x, y)$ are
measured simultaneously, as shown in Fig.~\ref{f:setup} (b) and (c)
respectively.  $\Delta f(x,y)$ is proportional to the force gradient
$\partial F_y / \partial y$ acting on the cantilever and is sensitive
to both the topography of the sample and to the magnetic field profile
in its vicinity.  Raised features such as the T-shaped top-electrode
of the nanoSQUID are visible.  $\Phi(x,y)$ shows a bipolar flux
response.  The change in sign of $\Phi(x,y)$ occurs as the Ni nanotube
crosses the $xz$-plane (defined by the SQUID loop) above the
nanoSQUID, matching the expected response.  Such images allow us to
identify the nanoSQUID and to position the Ni nanotube at a maximum of
$|\Phi(x, y)|$.  Given a constant $z$, the nanotube stray flux
optimally couples through the nanoSQUID loop at such positions,
resulting in the maximum signal-to-noise ratio for flux measurements.

\begin{figure}[b]
    \includegraphics[]{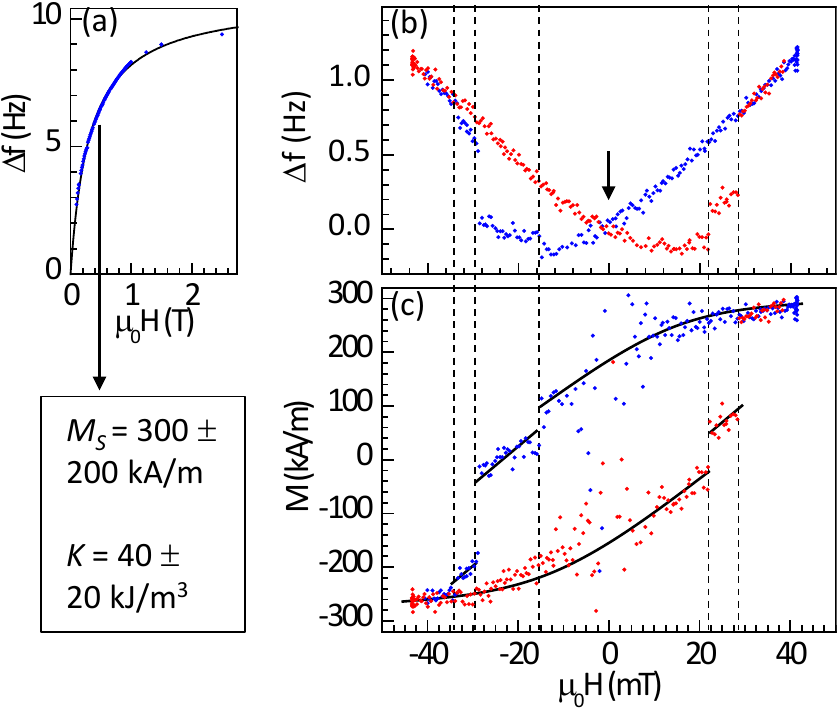}
    \caption{\label{f:cantileverHysteresis} (Color) (a) Cantilever
      magnetometry (points) and fit (solid line) in large magnetic
      fields.  (b) Cantilever magnetometry at small fields.  (c)
      Volume magnetization $M$ extracted from (b) according to
      (\ref{eq1}).  Solid lines guide the eye.  Red (blue) points
      represent data taken while sweeping $H$ in the positive
      (negative) direction.  Dashed lines highlight switching fields
      $H_{\rm sw,e}$. The error in $M$ scales with $1/|H|$, explaining
      the scatter near $H = 0$.}
\end{figure}

At one such position, indicated by the dot in Fig.~\ref{f:setup}, we
record $\Phi(H)$ by sweeping $\mu_0H$ from 41 mT to -41 mT and vice
versa.  A representative hysteresis curve $\Phi_{\text{NN}}(H) =
\Phi(H) - \Phi_{\text{ref}}(H)$ is shown in
Fig.~\ref{f:SQUIDHysteresis} (a) where $\Phi(H)$ is measured at
$z=450~$nm.  $\mu_0 |H|$ is incremented in steps of $0.2\,
\mathrm{mT}$ with a wait time of $1\, \mathrm{s}$ before each
acquisition.  The hysteresis has an almost square shape with a maximum
flux $\Phi_{\text{NN}}=75\, \mathrm{m\Phi_0}$ coupled into the
nanoSQUID.  The loop appears similar to stray-field hysteresis loops
obtained from a bistable Ni nanomagnet \cite{Meier:2000} and the Ni
nanowire of Ref.~\cite{Wernsdorfer:1996}, where $H$ was collinear with
the long axis.  Such a shape may suggest that at $H=0$ the remanent
magnetization $M_{\text R}\approx M_{\text S}$.  Increasing $H$ from
zero (see red branch in Fig.~\ref{f:SQUIDHysteresis} (a)), we first
observe a nearly constant flux, then a variation by about 30~\% along
with tiny jumps in a small field regime, and finally a large jump
occurring near 30~mT.  Similar to Ref.~\cite{Wernsdorfer:1996}, our
SQUID data suggest that almost all magnetic moments are reversed at
once near 30 mT via a large irreversible jump, i.e.\ via domain
nucleation and propagation.

We now turn to cantilever magnetometry, which is sensitive to $M(H)$.
$\Delta f$ is first measured simultaneously with $\Phi(H)$ at $z=450$
nm, as shown in Fig.~\ref{f:SQUIDHysteresis} (b).  The torque measured
via $\Delta f$ is found to exhibit tiny jumps and large abrupt changes
at exactly the same switching fields $H_{\rm sw,e}$ as
$\Phi_{\text{NN}}(H)$.  We note that switching fields vary from sweep
to sweep \cite{SupMat} as was observed in the Ni nanowire of
Ref.~\cite{Wernsdorfer:1996}; such behavior is expected if nucleation
is involved, given its stochastic nature.  Importantly, there is
always a one-to-one correspondence between switching fields observed
in $\Delta f$ and flux $\Phi_{\text{NN}}$ as highlighted by the dashed
lines in Fig.~\ref{f:SQUIDHysteresis}.  This correlation confirms that
the changes in $\Delta f$ and $\Phi_{\text{NN}}$ have a single origin:
the reversal of magnetic moments within the Ni nanotube.

In order to analyze $\Delta f(H)$ in terms of $M(H)$ it is important
to retract the Ni nanotube from the nanoSQUID by several $\mu$m.  We
therefore avoid magnetic interactions with both the diamagnetic
superconducting leads and the modulation current of the nanoSQUID.
These interactions lead to an enhanced $\Delta f$ and a branch
crossing (indicated by an arrow in Fig.~\ref{f:SQUIDHysteresis} (b))
occurring at finite $H$ rather than at $H=0$ as was reported in
Ref.~\cite{Ruffer:2012}.  After retracting the nanotube from the
nanoSQUID, we measure $\Delta f (H) = f_c (H) - f_0$ as shown in
Fig.~\ref{f:cantileverHysteresis}.  We start the acquisition at a
large positive field ($\mu_0 H =2.8~{\rm T}$), where the nanotube is
magnetized to saturation and then reduce $H$ to zero as shown in
Fig.~\ref{f:cantileverHysteresis} (a).  In large fields, the nanotube
behaves as a single-domain magnetic particle, i.e.\ it is magnetized
uniformly and $M$ rotates in unison as the cantilever oscillates in
the magnetic field.  Based on this assumption, we fit the results with
an analytical model for $\Delta f(H)$ \cite{Weber:2012}.  The volume
of the Ni nanotube $V_{\text{Ni}}$, $\omega_0$, and $k_0$ are set to
their measured values, while the saturation magnetization $M_\text{S}
= 300 \pm 200$ kA/m and the anisotropy parameter $K = 40 \pm 20$
kJ/m$^3$ are extracted as fit parameters.  The error in these
parameters is dominated by the error associated with the measurement
of the nanotube's exact geometry and therefore of $V_{\text{Ni}}$
\cite{SupMat}.  $M_\text{S}$ is consistent with the findings of
Ref.~\cite{Weber:2012} on similar nanotubes and with $406$ kA/m, known
as the saturation magnetization for bulk crystalline Ni at low
temperature \cite{Kittel:2005}.

Figure~\ref{f:cantileverHysteresis} (b) shows $\Delta f (H)$ taken in
the low-field regime.  In an opposing field, we observe discrete steps
in $\Delta f (H)$ indicating abrupt changes in the volume
magnetization $M$.  As expected the branch crossing (arrow) occurs at
$H=0$ and the overall behavior is consistent with measurements of
similar nanotubes \cite{Weber:2012}.  To analyze the low field data,
we adapt the analytical model to extract the dependence of the volume
magnetization $M$ on $H$, i.e.\ the field dependence of magnetization
averaged over the entire volume of the nanotube.  Solving the
equations of Ref.~\cite{Weber:2012} describing the frequency shift for
$M$, we find:
\begin{equation}
  M = \frac{2 k_0 l_e^2 K \Delta f}{H \left ( K V_{\text{Ni}} f_0 - k_0 l_e^2 \Delta f \right )},
  \label{eq1}
\end{equation}

\noindent where $l_e = 85$ $\mu$m is the effective cantilever length
for the fundamental mode.  $M(H)$ extracted from
Fig.~\ref{f:cantileverHysteresis} (b) is plotted in
Fig.~\ref{f:cantileverHysteresis} (c).  In both field sweep
directions, the magnetization is seen to first undergo a gradual
decrease as $|H|$ decreases.  Starting from $\sim300$ kA/m at +40 mT,
$M$ reduces to $\sim 200$ kA/m at 0~mT.  We find $M_{\text R}\approx
0.65~M_{\text S}$, in contrast with the SQUID data suggesting
$M_{\text R}\approx M_{\text S}$.  However, this gradual change of $M$
at small $|H|$ in the initial stage of the reversal is consistent with
the gradually changing anisotropic magnetoresistance observed in a
similar nanotube of larger diameter in nearly the same field regime
\cite{Ruffer:2012}.  At -15~mT, just before the first of three
discontinuous jumps, $M$ is only $\sim 100~$kA/m.  Note that jumps are
seen after the magnetization has decreased to a value of about
$0.3~M_{\text S}$.  Two further jumps occur at $\mu_0H_{\rm sw,e}=-28$
and -33~mT.  For $\mu_0H<-40$~mT, the nanotube magnetization is
completely reversed.  We observe a somewhat asymmetric behavior at
positive and negative fields.  This asymmetry may be due to an
anti-ferromagnetic NiO surface layer providing exchange interaction
with the Ni nanotube \cite{Zhang:2000,Gierlings:2002}.  Irreversible
jumps in $M$ are observed for $15~$mT~$<\mu_0|H_{\rm sw,e}|<35~$mT in
Fig.~\ref{f:cantileverHysteresis}, in perfect agreement with the range
over which jumps occur in $\Phi_{\rm NN}$ with the nanotube close to
the nanoSQUID in Fig.~\ref{f:SQUIDHysteresis}.

\begin{figure}[b]
  \includegraphics[width=8.5 cm]{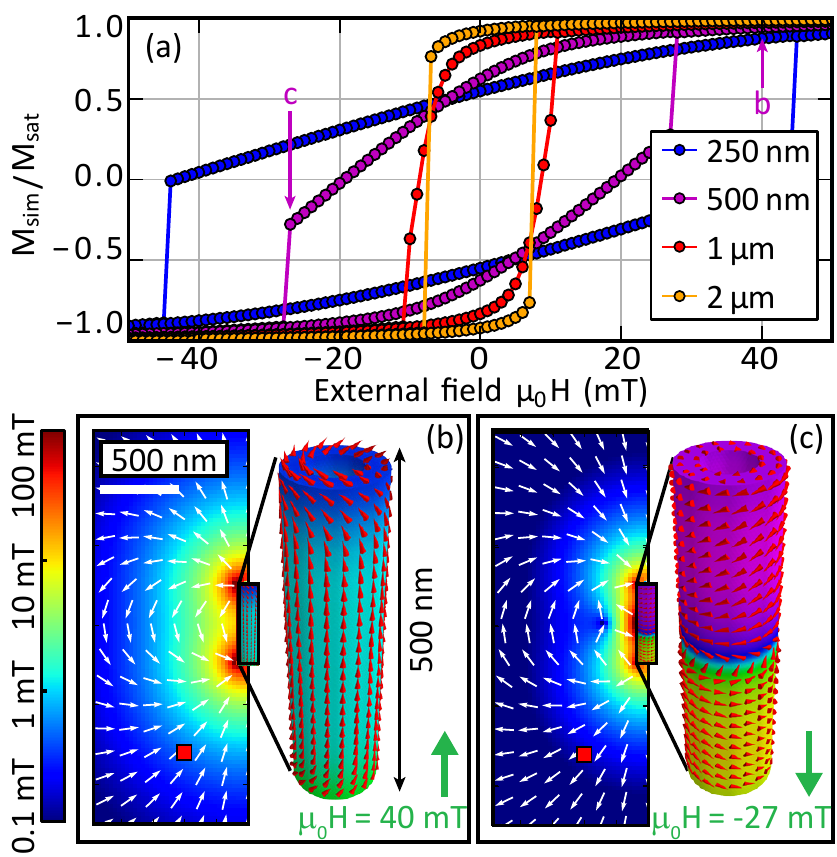}
  \caption{\label{f:simulation} (Color) (a) Simulated hysteresis loops
    $M(H)$ for nanotubes of four different $l$. $H_{\rm sw}$ increases
    with decreasing $l$.  Magnetic configurations (right) and
    stray-field distribution (left) for $l=500~$nm at (b) $40~$mT and
    (c) $-27~$mT as indicated by the labels in (a).  Cones (arrows)
    indicate the local direction of the magnetic moments (stray
    field).  The stray fields $H_{\rm str}$ are color coded as
    depicted.  The red squares indicate the position of the center of
    the nanoSQUID loop.}
\end{figure}

The observed magnetization steps suggest the presence of 2 to 4
intermediate magnetic states or 2 to 4 segments in the nanotube that
switch at different $H$.  Calculations for ideal nanotubes
\cite{Landeros:2009} suggest that the intermediate states should be
multi-domain, consisting of uniform axially saturated domains
separated by azimuthal or vortex-like domain walls.  The preferred
sites for domain nucleation are expected to be the two ends of the
nanotube \cite{Hertel:2004,Landeros:2009}.  As the field is reduced
after saturation, magnetic moments should gradually curl or tilt away
from the field direction.  The torque magnetometry measurements, which
show both gradual and abrupt changes in $M(H)$, are consistent with
such gradual tilting; the SQUID data, showing only abrupt changes in
$\Phi_{\text{NN}}(H)$, are not.  In the following we present
micromagnetic simulations performed on Ni nanotubes of different
lengths $l$ to further analyze our data.

In Fig.~\ref{f:simulation} (a) we show simulated hysteresis loops
$M(H)$ with $\mathbf{H}$ applied along the long axis of nanotubes with
$l$ between 250 nm and 2~$\mu$m.  For $l=2~\mu$m the $M(H)$ loop is
almost square, but the switching field is $\sim 8$~mT.  This value is
much smaller than the regime of $H_{\rm sw,e}$ observed
experimentally.  Nanotubes with 250~nm$~<l<1~\mu$m are consistent with
$15~{\rm mT}< \mu_0 |H_{\rm sw,e}|<35~$mT .  For $l=500~$nm the
simulation provides a switching field $\mu_0H_{\rm sw}=28~$mT.  At the
same time, $M$ is almost zero for $|H|$ just below $|H_{\rm sw}|$.
Such behavior is consistent with the overall shape of the measured
$M(H)$ loop in Fig.~\ref{f:cantileverHysteresis} (c), where the
largest jumps in $M$ take place at about $\pm 30~$mT.  Comparing
Fig.~\ref{f:simulation} (a) and Fig.~\ref{f:cantileverHysteresis} (c),
we conclude that the superposition of a few segments with $250~{\rm
  nm}<l<1~\mu$m could account for the measured $M(H)$.  For such
segments, Fig.~\ref{f:simulation} (b) and (c) (right panels) show
characteristic spin configurations (cones) well above and near $H_{\rm
  sw}$, respectively.  We observe the gradual tilting of spins at both
ends in (b) and two tubular-like vortex domains with opposite
circulation direction in (c) \cite{Chen:2011}.  Between the domains a
N{\'e}el-type wall exists.  For each $l$ and $M(H)$, we simulate the
relevant stray field at the position of the nanoSQUID (red squares in
the left panels of Fig. \ref{f:simulation} (b) and (c)) providing the
predicted $\Phi_{\rm NN}(H)$ \cite{SupMat}.  The shapes of the
simulated $\Phi_{\rm NN}(H)$ are nearly proportional to and thus
closely follow the shape of $M(H)$ shown in Fig.~\ref{f:simulation}
(a).  Thus the simulations allow us to explain the measured torque
magnetometry data, although they are inconsistent with the nanoSQUID
data.

The contrast between hysteresis traces obtained by the nanoSQUID and
torque magnetometry shows that $\Phi(H)$ is not the projection of
$\mathbf{M}$ along the nanotube axis.  This finding contradicts the
assumption of Ref.~\cite{Wernsdorfer:1996}; we attribute this
discrepancy to the fact that while cantilever magnetometry measures
the entire volume magnetization, the nanoSQUID is most sensitive to
the magnetization at the bottom end of the nanotube, as shown in
calculations of the coupling factor $\phi_\mu = \Phi/\mu$ (flux $\Phi$
coupled to nanoSQUID by a point-like particle with magnetic moment
$\mu$) \cite{Nagel:2013}.  Still, we find a one-to-one correspondence
between switching fields $H_{\rm sw,e}$ detected by either the
nanoSQUID or cantilever magnetometry.  This experimentally verified
consistency substantiates the reversal field analysis performed in
Ref.~\cite{Wernsdorfer:1996}.  In Fig.~\ref{f:SQUIDHysteresis} (a), we
find no clear evidence for curling or gradual tilting at small $H$.
The reversal process thus does not seem to start from the end closest
to the nanoSQUID, but rather from a remote segment.  This is an
important difference compared to the ideal nanotubes considered thus
far in the literature, in which both ends share the same fate in
initiating magnetization reversal.  The unintentional roughness of
real nanotubes might be relevant here.  In an experiment performed on
a large ensemble of nanotubes, one would have not been able to judge
whether a gradual decrease in $M(H)$ \cite{Albrecht:2011} originated
from a very broad switching field distribution or from the gradual
tilting of magnetic moments in the individual nanotubes.  Our
combination of nano-magnetometry techniques thus represents a powerful
method for unraveling hidden aspects of nanoscale reversal processes.
In order to optimize nanotubes for sensing and memory applications,
such understanding is critical.

In summary, we have presented a technique for measuring magnetic
hysteresis curves of nanometer-scale structures using a
piezo-electrically positioned nanoSQUID and a cantilever operated as a
torque magnetometer.  This dual functionality provides two independent
and complementary measurements: one of local stray magnetic flux and
the other of volume magnetization.  Using this method we gain
microscopic insight into the reversal mechanism of an individual Ni
nanotube, suggesting the formation of vortex-like tubular domains with
N{\'e}el type walls.

A. Buchter, J. Nagel, and D. R\"{u}ffer contributed equally to this
work which was funded by the Canton Aargau, the SNI, the SNF under
Grant No. 200020-140478, the NCCR QSIT, the DFG via the SFB/TRR 21,
and the ERC via the advanced grant SOCATHES.  Research leading to
these results was funded by the European Community's Seventh Framework
Programme (FP7/2007-2013) under Grant Agreement No. 228673 MAGNONICS.
J. Nagel acknowledges support from the Carl-Zeiss-Stiftung.

\end{document}